\documentclass[lettersize,journal]{IEEEtran}
\usepackage{amsmath,amsfonts}
\usepackage{algorithmic}
\usepackage{algorithm}
\usepackage{array}
\usepackage[caption=false,font=normalsize,labelfont=sf,textfont=sf]{subfig}
\usepackage{textcomp}
\usepackage{stfloats}
\usepackage{url}
\usepackage{verbatim}
\usepackage{graphicx}
\usepackage{cite}
\usepackage{multirow}
\usepackage{xcolor}
\usepackage[justification=centering]{caption}
\begin{document}



\title{Low-Light Image Enhancement Using Gamma Learning And Attention-Enabled Encoder-Decoder Networks}

\author{Bibhabasu Debnath, Sahana Ray, and Sanjay Ghosh,~\IEEEmembership{Senior Member,~IEEE,}
\thanks{This work is supported by the Faculty Start-up Research Grant (FSRG), Indian Institute of Technology Kharagpur, India.}
\thanks{Bibhabasu Debnath, Sahana Ray, and Sanjay Ghosh are with the Department of Electrical Engineering, Indian Institute of Technology Kharagpur, WB 721302, India (email: \texttt{ sanjay.ghosh@ee.iitkgp.ac.in}).}
}


\markboth{Under Review}%
{Shell \MakeLowercase{\textit{et al.}}: A Sample Article Using IEEEtran.cls for IEEE Journals}


\maketitle

\begin{abstract}
Images acquired in low-light environments present significant obstacles for computer vision systems and human perception, especially for applications requiring accurate object recognition and scene analysis. Such images typically manifest multiple quality issues: amplified noise, inadequate scene illumination, contrast reduction, color distortion, and loss of details. 
While recent deep learning methods have shown promise, developing simple and efficient frameworks that naturally integrate global illumination adjustment with local detail refinement continues to be an important objective. To this end, we introduce a dual-stage deep learning architecture that combines adaptive gamma correction with attention-enhanced refinement to address these fundamental limitations. The first stage uses an Adaptive Gamma Correction Module (AGCM) to learn suitable gamma values for each pixel based on both local and global cues, producing a brightened intermediate output. The second stage applies an encoder-decoder deep network with Convolutional Block Attention Modules (CBAM) to this brightened image, in order to restore finer details. We train the network using a composite loss that includes L1 reconstruction, SSIM, total variation, color constancy, and gamma regularization terms to balance pixel accuracy with visual quality. Experiments on LOL-v1, LOL-v2 real, and LOL-v2 synthetic datasets show our method reaches PSNR of upto 29.96 dB and upto 0.9458 SSIM, outperforming existing approaches. Additional tests on DICM, LIME, MEF, and NPE datasets using NIQE, BRISQUE, and UNIQUE metrics confirm better perceptual quality with fewer artifacts, achieving the best NIQE scores across all datasets. Our \texttt{GAtED} (\textbf{G}amma learned and \textbf{At}tention-enabled \textbf{E}ncoder-\textbf{D}ecoder) method  effectively handles both global illumination adjustment and local detail enhancement, offering a practical solution for low-light enhancement. The code should be accessible at \textcolor{blue}{https://github.com/bibhabasuiitkgp/GAtED}.
\end{abstract}



\section{Introduction}

Images captured in poor lighting conditions pose substantial challenges for both human observation and computer vision applications, particularly in object detection tasks~\cite{xu2021exploring, li2021photon, liu2021benchmarking}. When optical devices operate under inadequate illumination or when external lighting varies significantly, the resulting photographs exhibit multiple degradation artifacts: amplified noise, insufficient illumination, reduced contrast, color distortion, and loss of fine details. Manual correction of these images proves both time-intensive and frequently yields suboptimal results. Consequently, researchers have developed numerous low-light image enhancement (LLIE) algorithms to tackle these degradation problems and improve image quality.

Existing LLIE approaches fall into two broad categories: traditional methods and deep learning frameworks. Traditional methods build upon classical image processing principles such as histogram equalization~\cite{chen2003minimum, ibrahim2007brightness}, Retinex-based methods~\cite{ghosh2019fast} and optimization strategies~\cite{fu2016weighted, guo2016lime}. While~\cite{ibrahim2007brightness} improves contrast by adaptively redistributing pixel intensities,~\cite{guo2016lime} refine illumination maps through structure-aware enhancement. These conventional approaches offer computational efficiency and mathematical interpretability; however, they often suffer from color distortion and edge artifacts, and typically rely on restrictive assumptions about illumination distributions.

Deep learning-based approaches have achieved superior performance through data-driven learning. Within this category, Retinex-inspired models~\cite{wei2018deep, wu2022uretinex, zhang2019kindling, wu2025interpretable, liu2021retinex, xu2024cretinex, cai2023retinexformer, zheng2023empowering, yi2025diff, zhao2024riro} decompose images into illumination and reflectance components using trainable networks. KinD~\cite{zhang2019kindling} employs dedicated subnetworks for reflectance restoration and illumination enhancement, while Retinex-Net~\cite{wei2018deep} introduces trainable Decom-Net followed by encoder-decoder enhancement. Advanced frameworks like URetinex-Net++~\cite{wu2025interpretable} integrate implicit prior regularization and cross-stage fusion blocks for improved noise suppression and detail preservation. Retinexformer~\cite{cai2023retinexformer} leverages Illumination-Guided Transformers to model long-range dependencies and handle varying lighting conditions adaptively. CUE~\cite{zheng2023empowering} presents a Retinex-based Customized Unfolding Enhancer with Masked Autoencoder-based priors, while Diff-Retinex++~\cite{yi2025diff} integrates diffusion models with Retinex-driven restoration for physically-inspired generative enhancement. Among supervised CNN-based algorithms,~\cite{wang2020lightening} regards low-light enhancement as a residual learning problem, while ~\cite{shi2024ll} uses nested skip connections and residual blocks based on UNet++. On the other hand, unsupervised frameworks such as  Zero-DCE~\cite{guo2020zero} predicts high-order curves for pixel-wise adjustment without paired data, while EnlightenGAN~\cite{jiang2021enlightengan} leverages a lightweight GAN architecture to learn unpaired mappings from low- to normal-light images. More recently, diffusion-based frameworks~\cite{yi2025diff, yin2023cle, zhou2023pyramid, zheng2024selective, hou2023global, wang2024zero} have emerged as powerful approaches for image restoration and enhancement. PyDiff~\cite{zhou2023pyramid} employs progressive pyramid-resolution sampling strategies with global correctors for generalization to unseen noise distributions. DiffUIR~\cite{zheng2024selective} proposes selective hourglass mapping to handle diverse degradation distributions simultaneously. GSAD~\cite{hou2023global} introduces uncertainty-guided regularization for enhanced detail and noise suppression. Despite their generative capabilities, these models require significant computational resources and multiple sampling steps, limiting real-time applicability.
Contemporary LLIE research has increasingly incorporated attention mechanisms \cite{woo2018cbam} and adaptive gamma correction to handle non-uniform illumination and multi-scale degradations \cite{zamir2022restormer,wang2023ultra,yang2023implicit}. 
Yang et al.~\cite{yang2021low} proposed a Retinex-based framework incorporating pixel-wise gamma maps following image decomposition, learning adaptive corrections for uneven lighting while preserving structure. Wang et al.~\cite{wang2023low} introduced hierarchical gamma maps learned through Taylor expansion coupled with transformer-based global modeling, accelerating inference while maintaining adaptive illumination adjustment.

In this work, we propose a dual-stage enhancement architecture that leverages an Adaptive Gamma Correction Module (AGCM) followed by a U-Net augmented by a convolutional block attention module (CBAM). The first stage (AGCM) computes a gamma correction parameter map and applies a power law correction. The second stage (U-Net augmented CBAM) is a U-Net architecture integrated with Convolutional Block Attention Modules (CBAM) that learns to enhance contrast, restore the edge, and texture restoration, and noise suppression. The motivation behind the dual-stage enhancement architecture is that global illumination simplifies the task for the U-net to recover textures and colors.

The key contributions in our dual-stage method are:
\begin{itemize}
    \item Adaptive Gamma Correction Metwork (AGCM) for illumination recovery of the underexposed image by applying a differentiable learned gamma correction map. The gamma Correction map estimates pixel-wise gamma values to adaptively brighten different regions based on content. But while boosting brightness and improving visibility, it fails to restore the fine textures, Noise suppression, and color balancing. To address the above problem, the CBAM-enhanced U-Net architecture is introduced after the gamma-corrected sample.
    \item Attention-enhanced U-Net for local detail recovery and perceptual quality. This is designed to refine the gamma-corrected images by restoring lost textures, suppressing noise, and color imbalance. The architecture integrates Convolutional Block Attention Modules (CBAM) within a U-Net framework to dynamically focus on salient spatial regions and informative channels during both encoding and decoding. While the gamma-corrected image offers a globally brightened view, it often lacks structural sharpness and perceptual consistency. The attention-enhanced U-Net addresses these limitations by leveraging skip connections, multi-scale features, and attention-guided refinement to recover fine details and enhance overall image realism.
    \item Extensive experiments are performed on LOL-v1 and LOL-v2 (Real-Captured) datasets, comparing the proposed method with several of its state-of-the-art alternatives. Results demonstrate the superiority of our proposed \textbf{GAtED} (\textbf{G}amma Learned and \textbf{At}tention-enabled \textbf{E}ncoder-\textbf{D}ecoder) method.

\end{itemize}

The rest of this paper follows a systematic presentation of our contributions and their significance. In Section II, the existing literature on LLIE is explored, covering both traditional methods and contemporary deep learning approaches, with particular emphasis on attention mechanisms and adaptive gamma correction techniques. Section III details our proposed GAED methodology, including its architecture, mathematical formulations, and the expression of the multi-term loss function. Section IV outlines our experimental framework, introducing the datasets used, the metrics evaluated, and the implementation specifications. Additionally, it highlights the superior performance achieved by our proposed method compared to its state-of-the art alternatives, through qualitative and quantitative evaluations. Section V concludes with a discussion of the significance and achievements of the contributions proposed in this work.

\section{Related Work}

Over the past decades, low-light image enhancement (LLIE) has received significant research attention, resulting in a variety of traditional and deep learning-based approaches.

\subsection{Traditional LLIE Methods}

\subsubsection{Histogram-Based Methods}
Histogram Equalization (HE)~\cite{chen2003minimum} improves image contrast by redistributing pixel intensity values but often results in over-enhancement and noise amplification, particularly in extremely dark or bright regions.  
Dynamic Histogram Equalization (DHE)~\cite{ibrahim2007brightness} addresses this issue by dividing the histogram into sub-histograms based on local minima and allocating adaptive dynamic gray-level ranges, thus preserving local details and preventing dominant regions from overwhelming weaker features.

\subsubsection{Retinex-Based Methods}
Retinex theory~\cite{land1967lightness, edwin1977retinex} models an image as a product of illumination and reflectance. Since illumination is not directly observable, it must be estimated through various strategies, such as passing the image through filters~\cite{jobson1997properties, ghosh2019fast}.  
Single-Scale Retinex (SSR) and Multi-Scale Retinex (MSR)~\cite{elad2005retinex, jobson1997multiscale} enhance visibility by improving the illumination component using logarithmic transformations.  

\subsubsection{Optimization-Based Methods}
LIME~\cite{guo2016lime} estimates an initial illumination map by taking the pixel-wise maximum across RGB channels and refines it via an optimization method, producing structure-aware enhancement. LIME's use of the max-RGB heuristic may lead to inaccurate illumination estimation under colored lighting and introduces edge noise, limiting its robustness.

\subsection{Deep Learning-Based LLIE Methods}

\subsubsection{Retinex-Inspired Models}
KinD~\cite{zhang2019kindling} designs two dedicated subnetworks for reflectance restoration and illumination enhancement, helping preserve structural integrity and reduce color distortion.
Retinex-Net~\cite{wei2018deep} introduces a trainable Decom-Net to decompose the image into illumination and reflectance, followed by an Enhance-Net to boost illumination using an encoder-decoder structure.  
URetinex-Net~\cite{wu2022uretinex} unfolds an optimization-based Retinex formulation into a trainable network and integrates a learnable illumination adjustment module to better adapt to challenging lighting conditions.
URetinex-Net++~\cite{wu2025interpretable} further enhances this framework by introducing implicit prior regularization and a cross-stage fusion block, enabling adaptive decomposition with improved noise suppression, detail preservation, and color fidelity.
RUAS~\cite{liu2021retinex} builds a lightweight Retinex-inspired unrolled network and employs a cooperative reference-free learning strategy to discover effective low-light prior architectures, enabling fast and resource-efficient image enhancement.
CRetinex~\cite{xu2024cretinex} introduces a color-shift aware Retinex model that decomposes images into reflectance, color shift, and illumination components, enabling improved color constancy.
Retinexformer~\cite{cai2023retinexformer} addresses hidden corruptions in low-light images through a one-stage Retinex-based framework that employs an Illumination-Guided Transformer to model long-range dependencies and adaptively handle varying lighting conditions.
CUE~\cite{zheng2023empowering} presents a Retinex-based Customized Unfolding Enhancer that leverages Masked Autoencoder-based illumination and noise priors within structure and optimization flows, enabling more effective illumination restoration and noise suppression in low-light image enhancement.
Diff-Retinex++~\cite{yi2025diff} introduces a Retinex-driven reinforced diffusion model that integrates a diffusion model with Retinex-driven restoration, with the aim of achieving physically-inspired generative enhancement.
Retinex-Inspired Reconstruction Optimization (RIRO) model treats low-light enhancement as an image reconstruction task using spatial and frequency domain features~\cite{zhao2024riro}.


\subsubsection{CNN-Based Models}
In ~\cite{wang2020lightening}, low-light enhancement is treated as a residual learning problem and uses a Deep Lightening Network (DLN) to estimate the residual between normal-light and low-light images. On the other hand, ~\cite{shi2024ll} enhances low-light images using a UNet++ architecture with nested skip connections and Instance Normalization-based residual blocks.
%
Zero-DCE~\cite{guo2020zero} takes a different approach by predicting high-order curves for pixel-wise adjustment on the input image in an unsupervised manner using a Deep Curve Estimation Network.
EnlightenGAN~\cite{jiang2021enlightengan} leverages a lightweight, one-path GAN architecture to learn an unpaired mapping from low- to normal-light images without relying on paired data, enabling faster and more flexible training.

\subsubsection{Diffusion-Based Frameworks}
Diffusion models like ~\cite{yi2025diff} have recently emerged as a powerful approach to image restoration and enhancement. The Controllable Light Enhancement (CLE) Diffusion Model \cite{yin2023cle} introduces an illumination embedding for user-adjustable brightness and integrates the Segment-Anything Model (SAM) for user-defined region-specific enhancement. Pyramid Diffusion Model (PyDiff) employs a progressive pyramid-resolution sampling strategy and incorporates a global corrector to make the model faster and generalizable to unseen noise and illumination distributions~\cite{zhou2023pyramid}. DiffUIR framework~\cite{zheng2024selective} proposes a selective hourglass mapping strategy and shared distribution term (SDT), enabling the model to handle different degradation distributions at once. GSAD~\cite{hou2023global} introduces  uncertainty-guided regularization to enhance detail, contrast, and noise suppression in low-light images. In  ~\cite{wang2024zero}, a zero-reference low-light enhancement framework is proposed, leveraging physical quadruple priors and a pretrained generative diffusion model.

\subsubsection{Transformers and Attention Mechanisms}
The attention mechanism was initially proposed for visual processing~\cite{tsotsos1995modeling}. It allows neural networks to focus on important features while suppressing irrelevant information~\cite{mnih2014recurrent}.  
Dual attention modules like Convolutional Block Attention Module (CBAM)~\cite{woo2018cbam} combine both types sequentially. 
In LLIE, attention helps to identify underexposed regions, suppress noise, and improve structure. ~\cite{guo2020zero} and ~\cite{lv2021attention} used attention to enhance brightness and detail adaptively. Restormer~\cite{zamir2022restormer} redesigns the multi-head attention and feed-forward network blocks to capture long-range pixel interactions. 
LLFormer~\cite{wang2023ultra}  featured axis-based multi-head self-attention and cross-layer attention fusion. NeRCo~\cite{yang2023implicit} employs an implicit neural representation with semantic-oriented supervision to robustly recover perceptual-friendly visual results.



\subsubsection{Adaptive Gamma Correction-Based Methods:} 
%
Yang et al.~\cite{yang2021low} proposed a lightweight Retinex-based deep framework that incorporates an adaptive gamma correction module following image decomposition into illumination and reflectance components. 
Wang et al.~\cite{wang2023low} further introduced an illumination-aware gamma correction network, where hierarchical gamma maps are learned and approximated via Taylor expansion to accelerate inference, coupled with a transformer-based global modeling module. Unlike traditional approaches, these methods adaptively infer gamma values instead of using hand-crafted histograms or fixed curves.



\section{Proposed Method}
\begin{figure*}
\includegraphics[width=0.98\textwidth]{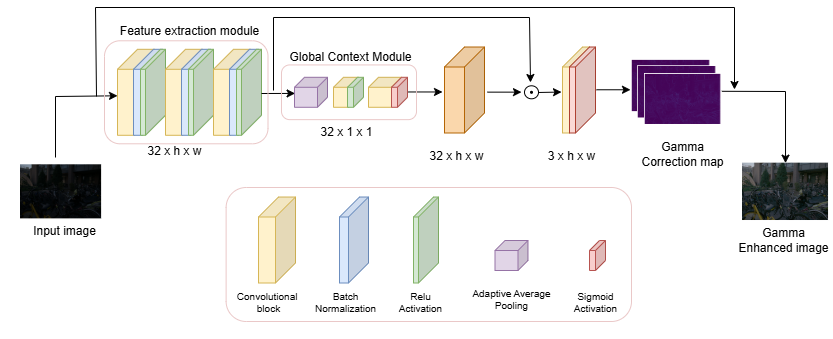}
\caption{Adaptive Gamma Correction Module (AGCM): the first stage of our proposed dual-stage deep learning method for low-light enhancement. It first extracts local textures and then injects global luminance cues to learn pixel-wise gamma for content-aware brightening. Note that we restricted the $\gamma$ values in $[0.5, 2.0]$ to avoid over-exposed while also preserving well-lit regions in the image.}
\label{fig:gamma}
\end{figure*}

\begin{figure*}
\centering
\includegraphics[width=0.95\textwidth]{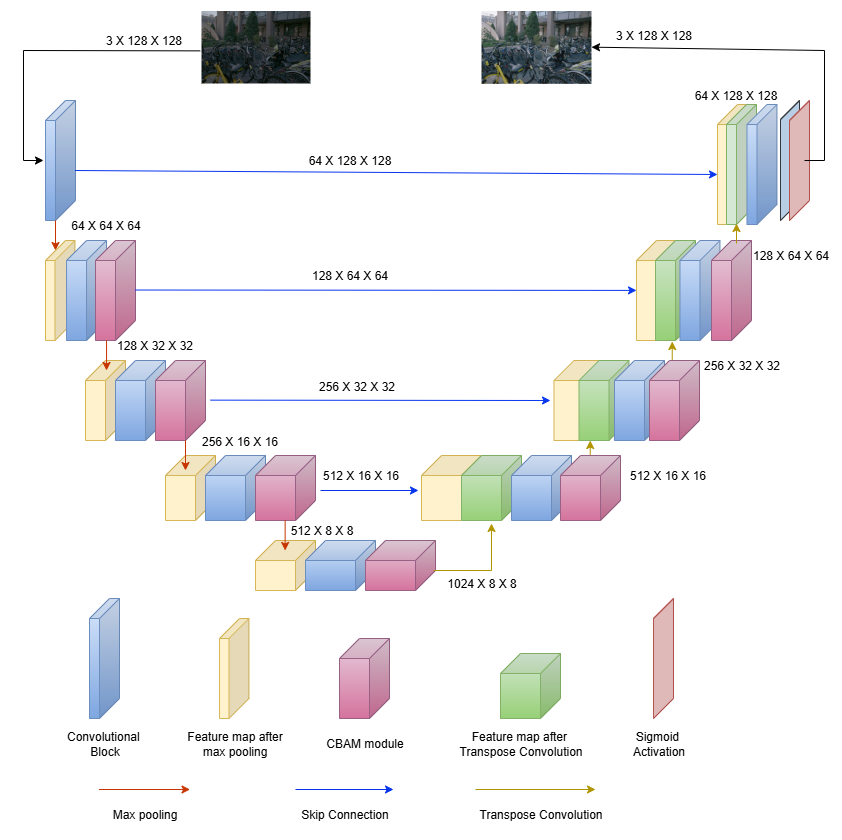}
\caption{CBAM‑enhanced U‑Net for refinement. Channel attention amplifies informative features, and spatial attention localises salient regions for enhancement. Skip connections with upsampling restore details while suppressing noise and colour imbalance. A final 1×1 convolution maps refined features to the enhanced image.}
\label{fig:UNet}
\end{figure*}

In this section, we present the complete details of our proposed deep learning method. We first introduce the proposed two-stage network architecture. This is followed by the details of our proposed loss function to guide the training of the proposed deep learning method. 
The first stage of the network, shown in Figure \ref{fig:gamma}, consists of an enhancement module that basically produces an enhanced image by performing a spatially adaptive gamma correction operation. In the second stage, we further enhance the low-light image by integrating a convolutional block attention module (CBAM) into an UNet framework, applying both channel and spatial attention sequentially, as shown in Figure \ref{fig:UNet}.



\subsection{Network Architecture}
\subsubsection{Adaptive Gamma Correction Module (AGCM)}

This module aims to produce an initial enhancement of the low-light image by adjusting illumination in a spatially adaptive manner. Traditional gamma correction applies a fixed exponent to all pixels, which fails to address non-uniform illlumination. To address this, our AGCM predicts a pixel-wise gamma map that adapts the gamma value based on both local texture and global context. The AGCM consists of a hierarchical \textit{Feature Extraction Block (FEB)} followed by a \textit{Global Context Block (GCB)}.

\paragraph{Feature Extraction Block (FEB)}  
The FEB captures local structures and textures of the input image. It uses three successive $3\times3$ convolution layers, each followed by batch normalization and ReLU activation, as shown in Figure~\ref{fig:gamma}.  Each convolution extracts feature responses at increasing levels of abstraction, i.e., edges, corners, and local intensity patterns.  For an input low-light image $I_{\text{in}} \in \mathbb{R}^{3 \times H \times W}$, we obtain the deep feature representation $F_3 \in \mathbb{R}^{32 \times H \times W}$, which encodes local information that determines how much each pixel should be enhanced.

\paragraph{Global Context Block (GCB)}  
While local features describe textures and edges, they lack awareness of the overall illumination and contrast distribution across the entire image. The GCB addresses this by integrating global context through an adaptive pooling and attention mechanism. First, adaptive average pooling condenses spatial information into a global descriptor $G \in \mathbb{R}^{32 \times 1 \times 1}$, representing the overall luminance statistics. Then, two successive $1 \times 1$ convolutions, interleaved with ReLU and Sigmoid activations, model non-linear relationships between channels:
\begin{equation}
C_1 = \text{ReLU}(\text{Conv}_{1\times1}^{16}(G)), \quad
C_2 = \sigma(\text{Conv}_{1\times1}^{32}(C_1)).
\end{equation}
Here, $C_2$ acts as a channel attention map indicating the importance of each feature channel in the local feature map $F_3$. Multiplying $C_2$ with $F_3$ modulates the response of each channel according to global illumination cues:
\begin{equation}
F_{\text{ctx}} = F_3 \odot C_2,
\end{equation}
where $\odot$ represents point-wise multiplication.
This step allows darker or underexposed regions to receive stronger activations, enabling more effective enhancement later.

\paragraph{Pixel-wise Gamma Prediction and Correction}  
After contextual refinement, the module predicts a \textbf{gamma correction map} that determines how much to brighten or darken each pixel. A $1\times1$ convolution followed by a scaled Sigmoid activation produces this gamma map:
\begin{equation}
\gamma = 0.5 + 1.5\, \sigma(\text{Conv}_{1\times1}^{3}(F_{\text{ctx}})),
\end{equation}
where $\sigma$ denotes the sigmoid activation function applied after a $1\times1$ convolution with 3 output channels on the context-enhanced features $F_{\text{ctx}}$, scaled and shifted to map outputs to the range $[0.5, 2.0]$.
The $\sigma$ nonlinearity operation ensures all predicted gamma values lie between 0 and 1, and the scaling extends this range to $[0.5, 2.0]$. Pixels with $\gamma < 1$ are brightened (since raising to a power less than 1 increases intensity), while those with $\gamma > 1$ are slightly darkened to prevent overexposure.
Finally, the enhanced image is computed by applying pixel-wise gamma correction:
\begin{equation}
I_{\gamma}(x, y) = (I_l(x, y) + \varepsilon)^{\gamma(x, y)}, \quad \varepsilon = 10^{-6},
\end{equation}
where $I_l(x,y)$ represents the low-light input image intensity at pixel location $(x,y)$, raised to the power of the predicted gamma value $\gamma(x,y)$ at that location, with a small constant $\varepsilon$ added to prevent numerical instability for near-zero pixel values.
This equation performs a content-aware illumination mapping - darker pixels are brightened more aggressively, while brighter regions are preserved. Through this adaptive mechanism, the AGCM produces a balanced, illumination-corrected version of the input image, serving as the initial enhanced output for the next stage.


\subsubsection{Attention-enabled Encoder--Decoder Architecture}

Although AGCM improves illumination adaptively, it may not fully restore lost details or textures, especially in very dark regions. Hence, the second stage leverages an attention-augmented UNet architecture to refine and reconstruct the image with better contrast, structural consistency, and color balance.

\paragraph{Convolutional Block Attention Module (CBAM)}  
To strengthen the model's ability to focus on relevant features, we integrate the Convolutional Block Attention Module (CBAM) into the UNet encoder and decoder blocks. CBAM sequentially applies channel attention and spatial attention, helping the network to understand what and where to improve.

\textit{Channel Attention:}  
Each feature map $F \in \mathbb{R}^{C \times H \times W}$ is summarized using global average pooling and global max pooling, producing two channel descriptors ($F_{\text{avg}}$ and $F_{\text{max}}$).  
These pass through a shared two-layer MLP to capture non-linear dependencies between channels. The resulting channel attention map $M_c$ highlights the most informative feature channels:
\begin{equation*}
M_{\text{c}} = \sigma(\text{MLP}(F_{\text{avg}}) + \text{MLP}(F_{\text{max}})),
\end{equation*}
where $F_{\text{avg}}$ and $F_{\text{max}}$ are the globally averaged and max-pooled channel descriptors, each processed through a shared multi-layer perceptron (MLP), summed element-wise, and passed through a sigmoid activation $\sigma$ to produce channel-wise attention weights.
Multiplying $M_c$ with $F$ amplifies globally useful features (e.g., edges and texture cues) and suppresses irrelevant noise.

\textit{Spatial Attention:}  
After channel attention, spatial attention determines which regions of the image are most important. It compresses the refined feature map $F'$ along the channel dimension using average and max pooling:
\begin{align*}
F'_{\text{avg}} &= \text{Mean}_{\text{chan}}(F'), \\
F'_{\text{max}} &= \text{Max}_{\text{chan}}(F'), \\
F_{\text{cat}} &= \text{Concat}(F'_{\text{avg}}, F'_{\text{max}}), \\
M_{\text{s}} &= \sigma(\text{Conv}_{7\times7}(F_{\text{cat}})),
\end{align*}
where $F'_{\text{avg}}$ and $F'_{\text{max}}$ represent the channel-wise mean and maximum operations applied to the channel-refined feature map $F'$, which are then concatenated along the channel dimension to form $F_{\text{cat}}$, followed by a $7\times7$ convolution and sigmoid activation to generate the spatial attention map $M_{\text{s}}$.
This map assigns higher weights to spatial regions that are more relevant for enhancement. The final refined output is obtained as:
\begin{equation}
F_{\text{enhanced}} = M_{\text{s}} \odot F',
\end{equation}
where $\odot$ denotes element-wise multiplication between the spatial attention map $M_{\text{s}}$ and the channel-refined features $F'$.
Together, CBAM ensures that each feature map is selectively refined in both the channel and spatial dimensions, helping the UNet focus its enhancement efforts where they matter most.

\paragraph{Encoder--Decoder Framework}  
The attention-augmented UNet consists of symmetric encoder and decoder paths, each built from Double Convolution Blocks. Each block contains two consecutive $3\times3$ convolutions with batch normalization and ReLU activations:
\begin{align*}
F_1 &= \text{ReLU}(\text{BN}(\text{Conv}_{3\times3}(x))), \\
F_2 &= \text{ReLU}(\text{BN}(\text{Conv}_{3\times3}(F_1))),
\end{align*}
where the input $x$ undergoes a $3\times3$ convolution, batch normalization (BN), and ReLU activation to produce $F_1$, which is then processed through another identical convolution-normalization-activation sequence to yield $F_2$.
The encoder progressively reduces spatial dimensions via max-pooling, capturing hierarchical features. Each downsampling step is wrapped in a CBAM to refine the feature map before deeper encoding:
\begin{equation*}
D_i = \text{DoubleConv}(\text{CBAM}(\text{MaxPool}(F_{i-1}))), 
\end{equation*}
where $\ i = {1,2,3,4}$  and the feature map $F_{i-1}$ from the previous layer is first max-pooled to reduce spatial resolution, then refined through CBAM, and finally processed through a double convolution block to produce the encoder output $D_i$.
At the bottleneck, the representation $F_B \in \mathbb{R}^{1024 \times 8 \times 8}$ encodes both global structure and fine details.

The decoder then gradually restores spatial resolution through transposed convolutions. Skip connections from corresponding encoder layers preserve fine details, while CBAM modules guide attention during upsampling:
\begin{equation*}
U_i = \text{DoubleConv}(\text{CBAM}(\text{Concat}(\text{UpConv}(U_{i+1}), D_i))), 
\end{equation*}
where $i = {3, 2, 1, 0}$ and the decoder output $U_{i+1}$ from the previous layer is upsampled via transposed convolution (UpConv), concatenated with the corresponding encoder feature map $D_i$ from the skip connection, refined through CBAM, and processed by a double convolution block to produce $U_i$.
Finally, a $1 \times 1$ convolution followed by a Sigmoid activation maps the decoder output to the final enhanced image:
\begin{equation}
\hat{x} = \sigma(\text{Conv}_{1\times1}(U_0)) \in \mathbb{R}^{3 \times 128 \times 128},
\end{equation}
where the final decoder output $U_0$ is passed through a $1\times1$ convolution to reduce channels to 3 (RGB), followed by sigmoid activation $\sigma$ to constrain pixel values to $[0,1]$, producing the enhanced image $\hat{x}$.
This output combines the global brightness correction from the first stage with fine-grained restoration and detail enhancement from the second stage, yielding a visually balanced, high-quality result.

\subsection{Loss Function}

\subsubsection{L1 Reconstruction Loss}
We employ the \textit{L1 Reconstruction Loss} to ensure that the enhanced image remains faithful to the ground truth at the pixel level. This loss penalizes the absolute differences between the predicted and target pixel intensities as follows.
%
%
\begin{equation}
\mathcal{L}_{\text{L1}} = \frac{1}{CHW}  \sum_{c=1}^C \sum_{i=1}^H \sum_{j=1}^W 
\left| \hat{x}_{c,i,j} - x_{c,i,j} \right|
\end{equation}

\subsubsection{SSIM Loss}
In order to preserve structural similarity and perceptual quality in the output image, we define the structural similarity index measure (SSIM) loss. 
%
The SSIM metric between the enhanced image and the ground truth image is computed as: 
\begin{equation}
\text{SSIM} (x, \hat{x})= 
\frac{(2\mu_1\mu_2 + C_1)(2\sigma_{12} + C_2)}
     {(\mu_1^2 + \mu_2^2 + C_1)(\sigma_1^2 + \sigma_2^2 + C_2)}
\end{equation}
where $\mu_1$ and $\mu_2$ are the pixel sample means of the images; $\sigma_1$ and $\sigma_2$ are the sample variances of the images; $\sigma_{12}$ the sample covariance.  Here, $C_1$ and $C_2$ are constants set to $0.01^2$ and $0.03^2$, respectively, to stabilize the division. Finally, the SSIM loss is 
\
\begin{equation*}
    \mathcal{L}_{\text{SSIM}} = 1 - \text{SSIM} (x, \hat{x}).
\end{equation*}


\subsubsection{Total Variation (TV) Loss}
We have a loss component based on the total variation (TV) of the enhanced/output image to penalize abrupt intensity changes between neighboring pixels, promoting spatial smoothness while preserving important edges. We compute TV loss as follows:
We compute TV by first summing the squared difference of pixel-intensities in the horizontal direction and in the vertical direction; followed by normalized by the number of elements ($CHW$).
%
%
In our experiments, we scale the final TV loss using a hyperparameter \( \lambda_{\text{TV}} \), which controls the strength of the regularization.



\subsubsection{Color Constancy Loss}
This loss component attempts to enforce consistent color balance across the RGB channels and suppress color distortions. In this way, we expect the means of the red, green, and blue channels to be close to each other in the enhanced image. 
Given an output image \( x \in \mathbb{R}^{N \times 3 \times H \times W} \), the mean of each channel is computed spatially across height and width. 
Let \( \mu_r, \mu_g, \mu_b \) denote the mean intensities of the red, green, and blue channels, respectively. The loss is defined as follows:
\begin{equation}
\mathcal{L}_{\text{color}} = \lambda_c \cdot 
\left[ (\mu_r - \mu_g)^2 + (\mu_r - \mu_b)^2 + (\mu_g - \mu_b)^2 \right]
\end{equation}
where \( \lambda_c \) is a scalar weight to control the influence of this term. The loss penalizes large deviations between color channels, promoting a natural-looking white balance. In our case \( \lambda_c \) is kept at 0.5.

\subsubsection{Gamma Regularization Loss}
To prevent instability due to extreme or unnatural gamma corrections, we employ a \textit{Gamma Regularization Loss}, which constrains the predicted gamma maps toward a desired average.
%
We first compute the global mean gamma value \( \bar{\gamma} \) across all channels and spatial locations. Next, we regularize it toward a predefined target value \( \gamma_t \) (in our case, 1.0). The loss is defined as follows:
\begin{equation}
\mathcal{L}_{\gamma} = \lambda_\gamma \cdot (\bar{\gamma} - \gamma_t)^2
\end{equation}
%
%
Here, \( \lambda_\gamma \) is a weighting coefficient that controls the influence of regularization. This loss penalizes deviations from the desired global gamma intensity, promoting a perceptually stable and consistent brightness correction across the image.

~\\
\noindent
\textbf{Dual-Stage Composite Loss Function:} \\
Our training objective is governed by a \textit{Dual-Stage Loss Function}, which supervises both the intermediate gamma-corrected output and the final enhanced image. This design enables stage-wise learning while prioritizing the final perceptual quality.
Let the outputs of the two enhancement stages be denoted by $x_t$ (gamma-enhanced output). Recall, final enhanced output and  the ground-truth images are referred as $\hat(x)$ and $x$ respectively. The dual-stage loss incorporates multiple components at each stage: pixel-level reconstruction (L1), structural similarity (SSIM), spatial smoothness (TV), and color constancy. Additionally, a gamma regularization term is applied to the predicted gamma maps in the first stage to stabilize gamma values.

The Stage-1 loss (applied to $x_{t}$) is defined as:
\begin{align}
\mathcal{L}_\text{stage1} =\ & 
\alpha \cdot \mathcal{L}_\text{L1}(x_t, x) + 
\beta \cdot \mathcal{L}_\text{SSIM}(x_t, x) \nonumber \\
& + \gamma \cdot \mathcal{L}_\text{TV}(x_t) +
\delta \cdot \mathcal{L}_\text{color}(x_t) +
\mathcal{L}_{\gamma}(\Gamma),
\end{align}
where  $\Gamma$ is the predicted gamma correction map.
The Stage-2 loss (applied to $\hat{x}$) is defined similarly, but without the gamma regularization:
\begin{align}
\mathcal{L}_\text{stage2} =\ & 
\alpha \cdot \mathcal{L}_\text{L1}(\hat{x}, x) + 
\beta \cdot \mathcal{L}_\text{SSIM}(\hat{x}, x) \nonumber \\
& + \gamma \cdot \mathcal{L}_\text{TV}(\hat{x}) +
\delta \cdot \mathcal{L}_\text{color}(\hat{x})
\end{align}
To prioritize the final stage while still guiding intermediate corrections, the total loss is computed as a weighted sum of both stages:
\[
\mathcal{L}_\text{total} = 0.3 \cdot \mathcal{L}_\text{stage1} + 0.7 \cdot \mathcal{L}_\text{stage2}
\]
Here, \( \alpha, \beta, \gamma, \delta \) are hyper-parameters that control the relative contributions of the loss components. In our implementation, we empirically set \( \alpha = 0.5 \), \( \beta = 0.2 \), \( \gamma = 0.2 \), and \( \delta = 0.1 \), balancing the trade-off between pixel-wise accuracy, perceptual similarity, spatial smoothness, and color consistency.

\newcommand{\up}{$\uparrow$}
\newcommand{\down}{$\downarrow$}

\section{Experiments}

\subsection{Experimental Settings}

\subsubsection{Dataset} 
For evaluation of our GAtED method, we conducted experiments on 3 widely used datasets with paired low-light and normal-light images: LOL-v1 ~\cite{wei2018deep} , LOL-v2 real ~\cite{yang2021sparse} and LOL-v2 synthetic ~\cite{yang2021sparse}.
The LOL-v1 dataset comprises 485 pairs of low-light and normal-light images for training and 15 pairs for testing. Each image has a spatial resolution of 600 × 400 pixels, and all samples were captured under real-world illumination conditions.
The LOL-v2 data set is divided into two subsets: LOL-v2 Real and LOL-v2 Synthetic (Syn). The LOL-v2 Real subset contains 689 training pairs and 100 testing pairs of real low-light and corresponding normal-light images. In contrast, the LOL-v2 Syn subset consists of 900 training pairs and 100 testing pairs, where the low-light images are synthetically generated from normal-light counterparts to simulate diverse lighting degradations. To enhance the robustness of the model under varying lighting conditions, we combined both subsets, resulting in a total of 2,074 training pairs and 215 testing pairs in order to compare with various no-reference image quality assessment metrics.

We further validate our model's generalizability by comparing against standard low-light datasets without ground-truth: DICM~\cite{lee2013contrast}, LIME~\cite{guo2016lime}, MEF~\cite{ma2015perceptual}, and NPE~\cite{wang2013naturalness}.
The datasets DICM contains 64 low-light images of mainly landscape scenes, LIME contains 10 images mainly focused on dark street landscapes, MEF has 17 images of dark indoor and building scenes, NPE has 8 low-light images covering mostly natural sceneries.

\begin{table*}
\centering
\caption{Quantitative Comparison of Our GAtED and Existing Low-Light Image Enhancement Methods \\ on the LOLv1~\cite{wei2018deep}, LOLv2 Real and LOLv2 Syn Datasets~\cite{yang2021sparse}.}
\label{tab:Lol}
\renewcommand{\arraystretch}{1.2}
\setlength{\tabcolsep}{4pt}
\scriptsize
\begin{tabular}{l|l|cccc|cccc|cccc}
\hline
\multirow{2}{*}{\textbf{Methods}} & \multirow{2}{*}{\textbf{Venue}} & \multicolumn{4}{c|}{\textbf{LOLv1 Dataset}} & \multicolumn{4}{c|}{\textbf{LOLv2 real Dataset}} & \multicolumn{4}{c}{\textbf{LOLv2 syn Dataset}} \\
\cline{3-14}
& & PSNR & SSIM & LPIPS & MAE & PSNR & SSIM & LPIPS & MAE & PSNR & SSIM & LPIPS & MAE \\
\hline
LIME~\cite{guo2016lime} & TIP'17 & 16.554 & 0.429 & 0.405 & 0.123 & 15.105 & 0.402 & 0.426 & 0.145 & 16.570 & 0.740 & 0.210 & 0.121 \\
RetinexNet~\cite{wei2018deep} & BMVC'18 & 17.558 & 0.651 & 0.379 & 0.117 & 16.097 & 0.401 & 0.543 & 0.131 & 17.137 & 0.762 & 0.255 & 0.117 \\
KinD~\cite{zhang2019kindling} & ACMMM'19 & 17.648 & 0.775 & 0.175 & 0.123 & 16.828 & 0.759 & 0.224 & 0.129 & 18.930 & 0.809 & 0.174 & 0.107 \\
RUAS~\cite{liu2021retinex} & CVPR'21 & 16.405 & 0.500 & 0.270 & 0.153 & 15.326 & 0.488 & 0.310 & 0.162 & 13.404 & 0.645 & 0.364 & 0.196 \\
Zero-DCE~\cite{guo2020zero} & CVPR'20 & 19.524 & 0.703 & 0.330 & 0.106 & 18.059 & 0.574 & 0.313 & 0.131 & 17.756 & 0.816 & 0.168 & 0.124 \\
EnlightenGAN~\cite{jiang2021enlightengan} & TIP'21 & 17.483 & 0.651 & 0.322 & 0.135 & 18.640 & 0.676 & 0.309 & 0.132 & 16.573 & 0.775 & 0.212 & 0.133 \\
Restormer~\cite{zamir2022restormer} & CVPR'22 & 22.156 & 0.817 & 0.151 & 0.078 & 21.236 & 0.820 & 0.191 & 0.082 & 25.105 & 0.925 & 0.066 & 0.057 \\
LLFormer~\cite{wang2023ultra} & AAAI'23 & 23.649 & 0.819 & 0.169 & 0.063 & 20.154 & 0.809 & 0.207 & 0.070 & 24.229 & 0.920 & 0.063 & 0.065 \\
Retinexformer~\cite{cai2023retinexformer} & ICCV'23 & 23.386 & 0.833 & 0.140 & 0.067 & 22.254 & 0.831 & 0.112 & 0.074 & 25.488 & 0.930 & 0.061 & 0.054 \\
NeRCo~\cite{yang2023implicit} & ICCV'23 & 22.946 & 0.786 & 0.146 & 0.069 & 18.490 & 0.633 & 0.414 & 0.111 & 19.510 & 0.763 & 0.231 & 0.095 \\
CUE~\cite{zheng2023empowering} & ICCV'23 & 21.670 & 0.775 & 0.224 & 0.079 & 18.053 & 0.753 & 0.347 & 0.129 & 22.209 & 0.877 & 0.117 & 0.072 \\
C-Retinex~\cite{xu2024cretinex} & IJCV'24 & 19.866 & 0.807 & 0.196 & 0.096 & 18.265 & 0.789 & 0.285 & 0.122 & 20.366 & 0.876 & 0.125 & 0.086 \\
CLE Diffusion~\cite{yin2023cle} & ACMMM'23 & 21.282 & 0.794 & 0.157 & 0.080 & 21.995 & 0.798 & 0.191 & 0.088 & 20.284 & 0.852 & 0.102 & 0.108 \\
QuadPrior~\cite{wang2024zero} & CVPR'24 & 18.771 & 0.781 & 0.209 & 0.112 & 20.471 & 0.811 & 0.198 & 0.081 & 16.102 & 0.758 & 0.251 & 0.146 \\
PyDiff~\cite{zhou2023pyramid} & arXiv'23 & 23.275 & 0.858 & 0.107 & 0.074 & 21.940 & 0.841 & 0.178 & 0.082 & 22.989 & 0.929 & 0.082 & 0.070 \\
GSAD~\cite{hou2023global} & NeurIPS'23 & 22.978 & 0.851 & 0.103 & 0.076 & 20.190 & 0.847 & 0.112 & 0.102 & 24.219 & 0.927 & 0.052 & 0.065 \\
DiffUIR~\cite{zheng2024selective} & CVPR'24 & 22.407 & 0.834 & 0.157 & 0.073 & 19.710 & 0.825 & 0.211 & 0.105 & 19.611 & 0.863 & 0.160 & 0.103 \\
Diff-Retinex++~\cite{yi2025diff} & TPAMI'25 & 24.667 & 0.867 & 0.101 & 0.062 & \textbf{23.413} & 0.872 & 0.134 & \textbf{0.064} & 26.058 & 0.944 & \textbf{0.041} & 0.054 \\
LL-UNet++~\cite{shi2024ll} & TCI'24 & 23.005 & 0.8682 & 0.104 & -- & -- & -- & -- & -- & 18.02 & 0.687 & 0.347 & --\\
RIRO~\cite{zhao2024riro} & TCI'24 & 23.01 & 0.897 & \textbf{0.072} & -- & -- & -- & -- & -- & -- & -- & -- & --\\
DLN~\cite{wang2020lightening} & TIP'20 & -- & -- & -- & -- & 21.946 & 0.807 & -- & -- & 23.829 & 0.912 & -- & --\\
URetinexNet~\cite{wu2022uretinex} & CVPR'22 & 21.32 & 0.834 & 1.22 & 0.067 & -- & -- & -- & -- & -- & -- & -- & --\\
ALL-E+~\cite{liang2025aesthetics} & TPAMI'25 & 19.56 & 0.773 & 0.209 & -- & -- & -- & -- & -- & -- & -- & -- & --\\
URetinex-Net++~\cite{wu2025interpretable} & TPAMI'25 & 23.826 & 0.8411 & 0.2311 & 0.0589 & -- & -- & -- & -- & -- & -- & -- & --\\
\hline
\textbf{GAtED} & \textbf{Proposed} & \textbf{25.46} & \textbf{0.9140} & 0.088 & \textbf{0.0467} & 22.50 & \textbf{0.9047} & \textbf{0.0763} & 0.0762 & \textbf{29.96} & \textbf{0.9458} & 0.0515 & \textbf{0.0324} \\
\hline
\end{tabular}
\end{table*}

\begin{table*}
\centering
\caption{Quantitative comparison of GAtED and denoising results on DICM, LIME, MEF, and NPE \\ datasets, including NIQE$\downarrow$, BRISQUE (BR.)$\downarrow$, and UNIQUE (UN.)$\uparrow$.}
\renewcommand{\arraystretch}{1.2}
\setlength{\tabcolsep}{5pt}
\begin{tabular}{l l | c c c | c c c | c c c | c c c}
\hline
\multirow{2}{*}{Methods} & \multirow{2}{*}{Venue} &
\multicolumn{3}{c|}{DICM~\cite{lee2013contrast}} & \multicolumn{3}{c|}{LIME~\cite{guo2016lime}} & \multicolumn{3}{c|}{MEF~\cite{ma2015perceptual}} & \multicolumn{3}{c}{NPE~\cite{wang2013naturalness}} \\
 &  & NIQE & BR. & UN. & NIQE & BR. & UN. & NIQE & BR. & UN. & NIQE & BR. & UN. \\
\hline
LIME & TIP’17 & 3.75 & 24.99 & 0.78 & 3.85 & 18.65 & 0.53 & 3.65 & 18.10 & 0.65 & 4.44 & 17.73 & 0.93 \\
Retinex-Net & BMVC’18 & 4.47 & 30.38 & 0.75 & 4.60 & 26.42 & 0.52 & 4.41 & 21.90 & 0.97 & 4.60 & 22.59 & 0.81 \\
ISSR & ACMMM’20 & 4.14 & 26.93 & 0.59 & 4.17 & 18.82 & 0.83 & 4.22 & 25.84 & 0.87 & 4.02 & 21.27 & 0.99 \\
MIRNet & ECCV’20 & 5.15 & 37.56 & -0.19 & 5.59 & 37.93 & 0.31 & 4.38 & 33.81 & 0.48 & 5.41 & 31.05 & 0.16 \\
URetinex-Net & CVPR’22 & 3.95 & 25.24 & 0.85 & 4.34 & 26.48 & 0.93 & 3.79 & 21.08 & 1.18 & 4.69 & 27.84 & 0.99 \\
SNR & CVPR’22 & 4.62 & 31.05 & 0.12 & 4.49 & 31.65 & 0.53 & 4.09 & 27.18 & 0.53 & 4.36 & 22.59 & 0.36 \\
LLFlow & AAAI’22 & 3.73 & 24.29 & \textbf{1.02} & 3.82 & 26.53 & \textbf{1.03} & 3.84 & 25.67 & \textbf{1.39} & 4.23 & 16.78 & \textbf{1.28} \\
Zero-DCE & CVPR’20 & 3.56 & 25.52 & 0.82 & 3.77 & 21.10 & 0.73 & 3.28 & 19.83 & 1.22 & 3.95 & 19.28 & 1.07 \\
RUAS & CVPR’21 & 5.21 & 32.58 & -0.17 & 4.26 & 22.35 & 0.34 & 3.83 & 18.82 & 0.73 & 5.14 & 22.85 & 0.37 \\
ReLLIE & ACMMM’21 & 4.44 & 27.00 & 0.41 & 5.22 & 20.27 & 0.52 & 5.22 & 27.74 & 1.07 & 4.47 & 28.68 & 0.21 \\
SCL-LLE & AAAI’22 & 3.51 & 24.63 & 0.87 & 4.01 & 16.99 & 0.76 & 3.31 & 16.18 & 1.25 & 4.47 & 20.51 & 0.38 \\
SCI & CVPR’22 & 4.11 & 26.83 & 0.39 & 4.21 & \textbf{14.78} & 0.62 & 4.09 & 14.76 & 0.65 & 4.28 & 25.93 & 0.94 \\
PairLIE & CVPR’23 & 4.08 & 31.80 & 0.67 & 4.52 & 22.59 & 0.78 & 4.10 & 29.56 & 1.08 & 4.19 & 23.12 & -0.01 \\
NeRCo & ICCV’23 & 4.37 & 32.24 & 0.04 & 4.86 & 22.65 & 0.16 & 4.06 & 30.70 & 1.06 & 4.14 & 15.19 & 0.96 \\

ALL-E & TPAMI & 3.49 & 24.56 & 0.88 & 3.78 & 21.48 & 0.80 & 3.32 & 30.46 & 1.27 & 3.85 & 18.58 & 1.10 \\
ALL-E+NIMA & TPAMI & 3.47 & 24.99 & 0.87 & 3.77 & 21.42 & 0.69 & 3.32 & 30.27 & 1.28 & 4.16 & \textbf{14.45} & 0.94 \\
ALL-E+TANet & TPAMI & 3.81 & \textbf{21.43} & 0.87 & 3.84 & 18.52 & 0.61 & 3.26 & \textbf{13.16} & 1.29 & 4.14 & 15.19 & 0.96 \\
\hline
\textbf{GAtED} & Proposed & \textbf{2.96} & 25.49 & 0.771 & \textbf{3.76} & 22.47 & 0.82 & \textbf{2.76} & 23.58 & 0.76 & \textbf{3.09} & 18.36 & 1.01 \\
\hline
\end{tabular}
\label{tab:NIQE}
\end{table*}
%



\subsubsection{Metrics} For evaluation of the model performance we used metrics - PSNR (Peak Signal to Noise Ratio), SSIM (Stuctural Similarity Index Measure) ~\cite{wang2004image}, LPIPS (Learned Perceptual Image Patch Similarity) ~\cite{zhang2018unreasonable}, MAE (Mean Absolute Error)
Peak Signal-to-Noise Ratio (PSNR) quantifies the relationship between the maximum possible signal power of an image and the power of the noise that affects its fidelity. Structural Similarity Index (SSIM) evaluates the perceptual similarity between the enhanced image and its reference image based on luminance, contrast, and structure. Learned Perceptual Image Patch Similarity (LPIPS) measures perceptual differences between images using deep feature representations. The mean absolute error (MAE) computes the average absolute difference between the enhanced and reference images.

We also employ several no-reference image quality assessment (NR-IQA) metrics, including the Natural Image Quality Evaluator (NIQE)~\cite{mittal2012making}, which quantifies image quality based on deviations from natural scene statistics without requiring any training on human-rated data; the Unified No-reference Image Quality and Uncertainty Evaluator (UNIQUE)~\cite{zhang2021uncertainty}, a deep learning–based model designed to assess image quality across both synthetic and realistic distortion scenarios; and the Blind/Referenceless Image Spatial Quality Evaluator (BRISQUE)~\cite{mittal2012no}, which measures perceptual quality degradation by analyzing locally normalized luminance statistics in the spatial domain.

\subsubsection{Setup details} Our method is implemented using Pytorch and NVIDIA-RTX 3050 Ti GPU. During training we have used 100 epochs and all the images are resized to 128 X 128 pixels followed by normalization of the pixels to 0 to 1 range. For the backbone model of LPIPS we have use AlexNet.

\begin{figure*}
    \centering
    \setlength{\tabcolsep}{2pt} 
    \renewcommand{\arraystretch}{1.2} 

    \begin{tabular}{cccc}
        \textbf{Input} & \textbf{ZeroDCE} & \textbf{KinD} & \textbf{RetinexNet} \\
        \includegraphics[width=0.23\linewidth]{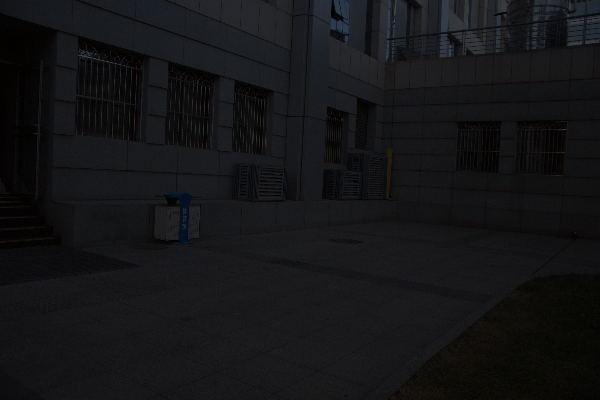} &
        \includegraphics[width=0.23\linewidth]{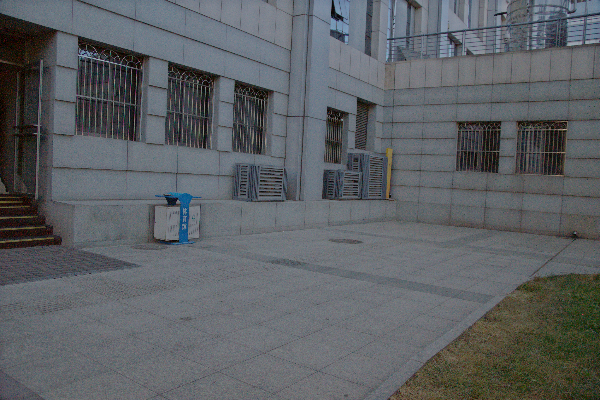} &
        \includegraphics[width=0.23\linewidth]{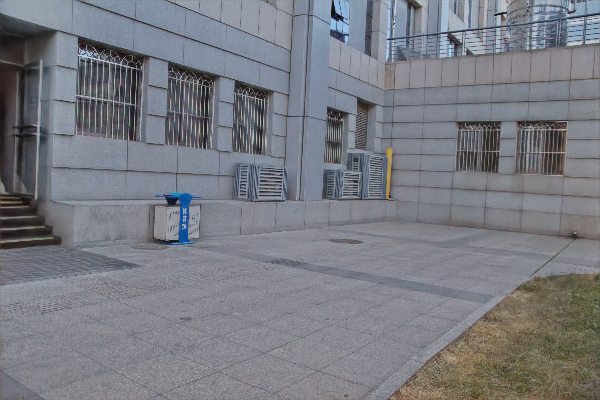} &
        \includegraphics[width=0.23\linewidth]{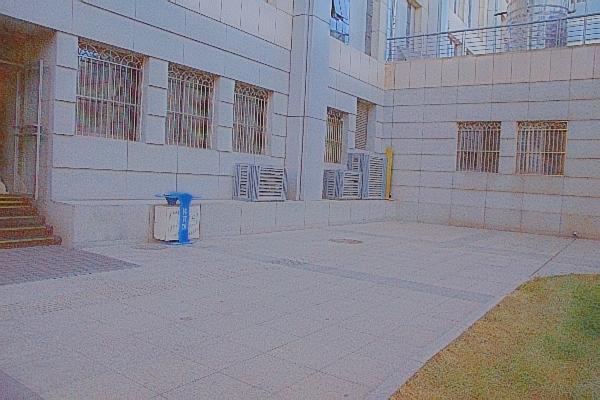} \\[4pt]

        \textbf{URetinexNet} & \textbf{NeRCo} & \textbf{GAtED} (Proposed) & \textbf{Ground-truth} \\
        \includegraphics[width=0.23\linewidth]{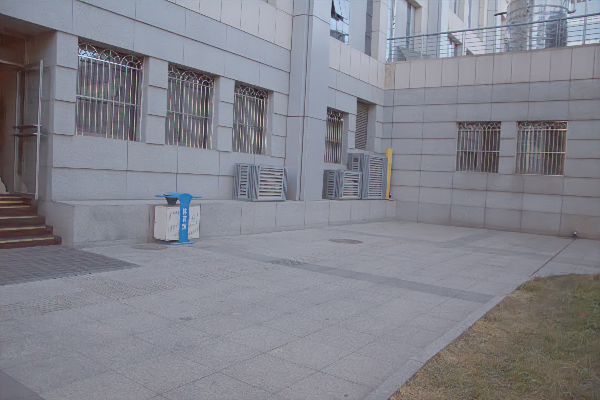} &
        \includegraphics[width=0.23\linewidth]{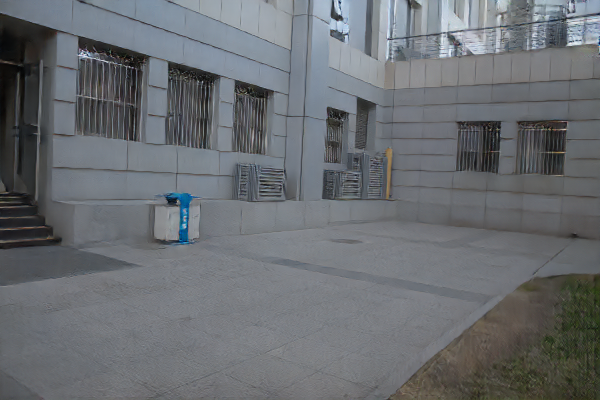} &
        \includegraphics[width=0.23\linewidth]{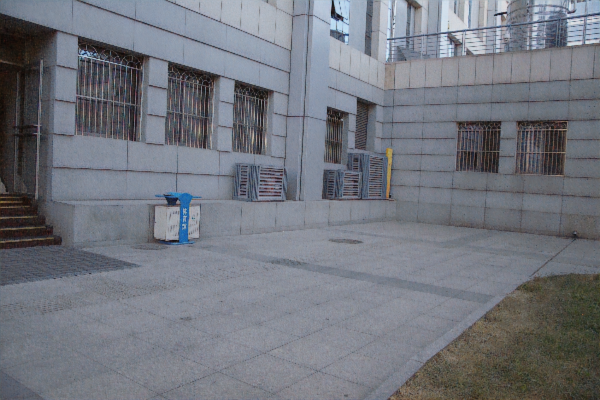} &
        \includegraphics[width=0.23\linewidth]{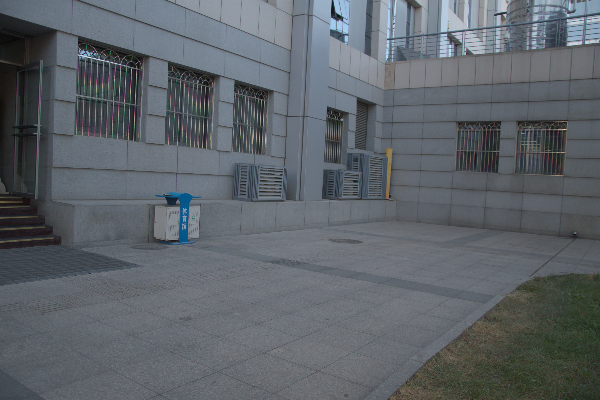} \\
    \end{tabular}
     \captionsetup{justification=centering}
    \caption{Visual comparison of different low-light image enhancement methods under extreme low light condition. Notice the appearance of unwanted color distortion in the output images of most of the methods compared. Unlike RetinexNet which tends to produce overexposed regions in brightly lit areas, our method GAtED produced content-aware illumination adjustment that preserves the natural luminance distribution across different image regions.}
    \label{fig:night}
\end{figure*}

\subsection{Quantitative Comparison with Benchmarking  Methods}
To thoroughly evaluate the effectiveness of our proposed approach, we conduct comprehensive comparisons with a diverse set of state-of-the-art low-light image enhancement methods across multiple benchmarks and evaluation metrics. The methods includes Retinex based methods LIME~\cite{guo2016lime}, RetinexNet~\cite{wei2018deep}, KinD~\cite{zhang2019kindling}, RUAS~\cite{liu2021retinex}, CUE~\cite{zheng2023empowering}, RetinexFormer~\cite{cai2023retinexformer}, C-Retinex~\cite{xu2024cretinex}, U-RetinexNet~\cite{wu2022uretinex}, Diff-Retinex++~\cite{yi2025diff}, U-RetinexNet++~\cite{wu2025interpretable} followed by diffusion based methods - CLE Diffusion~\cite{yin2023cle}, PyDiff~\cite{zhou2023pyramid}, GSAD~\cite{hou2023global}, Quad Prior~\cite{wang2024zero}, DiffUIR~\cite{zheng2024selective}, and other image enhancement methods ZeroDCE~\cite{guo2020zero}, EnlightenGAN~\cite{jiang2021enlightengan}, Restormer~\cite{zamir2022restormer}, NeRCo~\cite{yang2023implicit}, Restormer~\cite{zamir2022restormer}, LLFormer~\cite{wang2023ultra}, LL-UNet++~\cite{shi2024ll}, RIRO~\cite{zhao2024riro}, DLN~\cite{wang2020lightening}, ALL-E+~\cite{liang2025aesthetics}, ISSR~\cite{fan2020integrating}, MIRNet~\cite{zamir2020learning}, SNR~\cite{xu2022snr}, LLFlow~\cite{wang2022low}, ReLLIE~\cite{zhang2021rellie}, SCL-LLE~\cite{liang2022semantically}, SCI~\cite{ma2022toward}, PairLIE~\cite{fu2023learning}

\subsubsection{With availability of ground-truth images}

We conducted experiments on the LOL dataset ~\cite{wei2018deep} ~\cite{yang2021sparse}, which are standard benchmarks in the literature on low-light enhancement. The results are shown in Table \ref{tab:Lol}. We have reported four metrics (PSNR, SSIM, LPIPS, and MAE) to quantifying the similarity between the enhanced and ground-truth images. Our method consistently demonstrates competitive or superior performance across most metrics. 
In particular, our method GAtED surpasses all the metrics except LPIPS for LOL-V1 dataset, PSNR and MAE for LOL-v2 real dataset and LPIPS for LOL-v2 Synthetic dataset. It is important to note that our method achieves remarkably high SSIM values for all variants of LOL dataset. 

\subsubsection{Without availability of ground-truth images}

Furthermore, to assess perceptual quality in the absence of ground truth, evaluations were conducted on standard no-reference low-light datasets, including DICM~\cite{lee2013contrast}, LIME~\cite{guo2016lime}, MEF~\cite{ma2015perceptual}, and NPE~\cite{wang2013naturalness}.. The performance was quantified using NIQE, BRISQUE, and UNIQUE metrics, which provide a comprehensive measure of visual quality. As shown in Table \ref{tab:NIQE}, GAtED achieves the best NIQE scores across all datasets, indicating that it produces images with perceptual characteristics most closely aligned with natural scene statistics. 
Our outstanding NIQE performance across no-reference datasets (DICM: 2.96, LIME: 3.76, MEF: 2.76, NPE: 3.09) demonstrates the natural appearance and artifact-free quality of our enhanced images. 
This superior NIQE performance reflects the model’s ability to effectively enhance illumination while preserving structural fidelity and reducing unnatural artifacts, thereby yielding visually pleasing and realistic results.

Overall, our proposed model achieves consistently strong performance across all evaluated datasets (both types: with and without ground-truth images) and metrics. This demonstrates its capability to generalize well to varying lighting conditions and its effectiveness in both pixel-based and perceptual quality enhancement.

\subsection{Visual Comparision}

\begin{figure*}
    \centering
    \setlength{\tabcolsep}{2pt} 
    \renewcommand{\arraystretch}{1.2} 

    \begin{tabular}{cccc}
        \textbf{Input} & \textbf{ZeroDCE} & \textbf{KinD} & \textbf{RetinexNet} \\
        \includegraphics[width=0.23\linewidth]{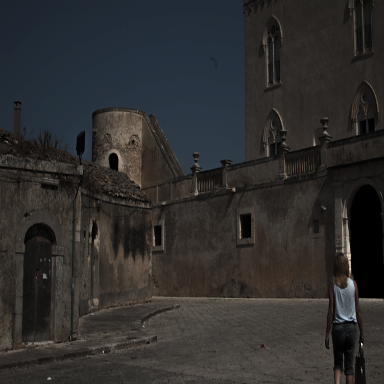} &
        \includegraphics[width=0.23\linewidth]{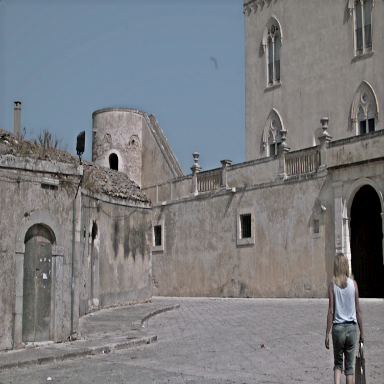} &
        \includegraphics[width=0.23\linewidth]{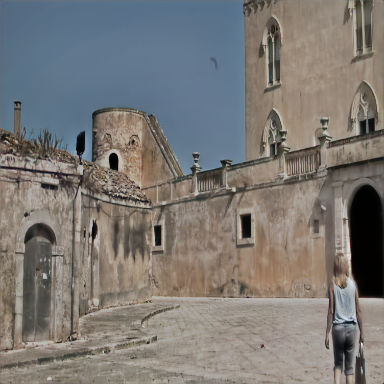} &
        \includegraphics[width=0.23\linewidth]{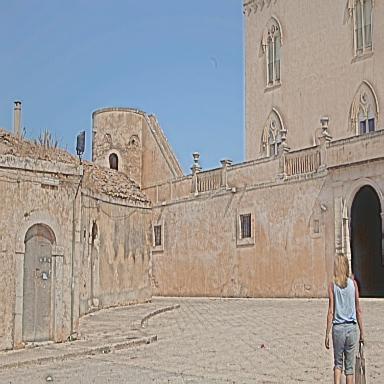} \\[4pt]

        \textbf{URetinexNet} & \textbf{NeRCo} & \textbf{GAtED} (Proposed) & \textbf{Ground-truth} \\
        \includegraphics[width=0.23\linewidth]{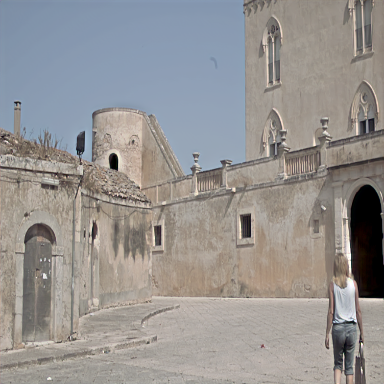} &
        \includegraphics[width=0.23\linewidth]{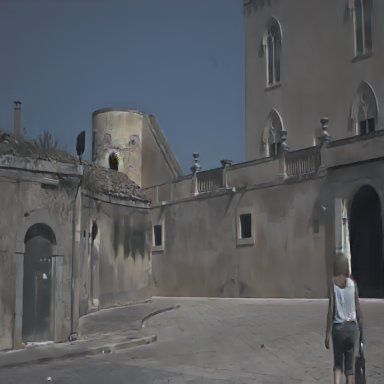} &
        \includegraphics[width=0.23\linewidth]{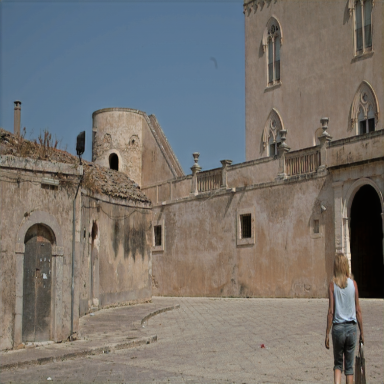} &
        \includegraphics[width=0.23\linewidth]{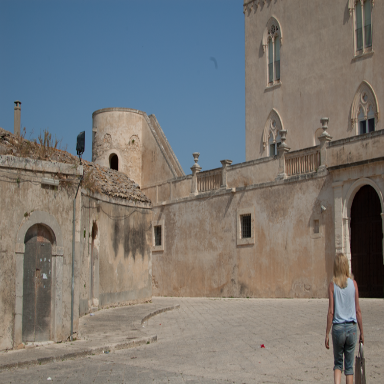} \\
    \end{tabular}
     \captionsetup{justification=centering}
    \caption{Visual comparison of different low-light image enhancement methods under moderate darkness. It is visually evident that our GAtED method consistently achieved a superior brightness enhancement while maintaining a natural appearance.}
    \label{fig:dawn}
\end{figure*}

To demonstrate the effectiveness of our proposed method, we visually compared it with some of the state-of-the-art methods, KinD~\cite{zhang2019kindling}, 
ZeroDCE~\cite{guo2020zero}, RetinexNet~\cite{wei2018deep}, 
U-RetinexNet~\cite{wu2022uretinex}, NeRCo~\cite{yang2023implicit}, in various challenging scenarios. 
We present visual comparisons on two low-light images (with varying low-lightness) in Figures~\ref{fig:night} and \ref{fig:dawn}.

The night scene (input image) in Figure~\ref{fig:night} is captured under extreme low-lighting conditions. Although it is hardly visible in the low-light image, the ground-truth image confirms the presence of fine details in the green grass region. The output images of most of the methods compared failed to successfully preserve fine structures and textures without color distortion. 
Unlike RetinexNet which tends to produce overexposed regions in brightly lit areas, or KinD which often results in underexposed dark regions, our method GAtED provides content-aware illumination adjustment that preserves the natural luminance distribution across different image regions.

In Figure \ref{fig:dawn}, we display the results on an image captured in the early morning. Although it has a relatively better lighting condition than the previous low-light image, this image is rich in strong structures. With close visual comparison with reference to the ground-truth, it is evident that most of the existing methods (shown here) suffer from color distortion. In particular, there is a prominent lack of brightness in the outputs of ZeroDCE and NeRCo, whereas KinD also has a slight drop in  brightness. The outputs of both RetinexNet and URetinexNet exhibit over-enhancement. In fact, RetinexNet results in significant structural distortion. 
In contrast, our GAtED method consistently achieves a superior brightness enhancement while maintaining a natural appearance.

In summary, both visual evaluations encompass both the improvement of overall image quality  and detailed preservation of fine structures and textures. A critical advantage of our method is the preservation of color consistency and the prevention of color distortion artifacts. As shown in the comparison images, methods like RetinexNet and NeRCo exhibit noticeable color shifts, particularly in regions with complex lighting conditions. Our attention-enabled encoder decoder network effectively maintains color balance through its attention-guided refinement process, resulting in enhanced images that closely match the ground truth color characteristics. 
The visual comparisons on these datasets show that our method produces images with characteristics most closely aligned with natural scene statistics, avoiding the over-enhancement and artificial appearance that plague other methods



%

\section{Conclusion}

In this work, we proposed a novel two-stage deep learning method for low-light image enhancement. 
%
that effectively integrates adaptive illumination correction and attention-guided refinement. The core idea was to first perform a learnable gamma correction which was followed by incorporating an attention mechanism in an encoder-decoder deep network. 
The first deep stage adaptively learned content-aware gamma maps to recover overall visibility, while the attention-enabled encoder-decoder network recovered structural detail, suppressed noise, and corrected color imbalances.
Experimental results confirmed that our model could address both global brightness inconsistencies and local perceptual degradations.
It consistently achieved competitive performance on multiple image quality metrics (PSNR, SSIM, LPIPS, MAE, NIQE, BRISQUE, and UNIQUE) compared to existing methods.

In future work, we plan to extend the proposed method for enhancing low-light videos. The process of video capturing in real often leads to blurring (for example, hand-shaking blur, motion blur) of the video frames. Therefore, we aim to solve the joint deblurring and enhancement of low-light image and video. We also plan to translate our method into practical applications, such as mobile
phones and embedded devices, to make it suited to a wide range of applications such as surveillance and autonomous vehicles.




\bibliographystyle{IEEEtran}
\bibliography{ref}

\end{document}